\documentstyle[12pt,epsf,epsfig]{article}
\def\beq{\begin{equation}}
\def\eeq{\end{equation}}
\def\bea{\begin{eqnarray}}
\def\eea{\end{eqnarray}}
\def\bq{\begin{quote}}
\def\eq{\end{quote}}

\def\gappeq{\mathrel{\rlap {\raise.5ex\hbox{$>$}}
{\lower.5ex\hbox{$\sim$}}}}

\def\lappeq{\mathrel{\rlap{\raise.5ex\hbox{$<$}}
{\lower.5ex\hbox{$\sim$}}}}

\def\laq{\ \raise 0.4ex\hbox{$<$}\kern -0.8em\lower 0.62
ex\hbox{$\sim$}\ }
\def\gaq{\ \raise 0.4ex\hbox{$>$}\kern -0.7em\lower 0.62
ex\hbox{$\sim$}\ }
\def\half{\hbox{\magstep{-1}$\frac{1}{2}$}}

\begin{document}
\pagestyle{empty}
\begin{flushright}
{CERN-TH/99-22}\\
{hep-th/9902097}\\
\end{flushright}
\vspace*{5mm}
\begin{center}
{\bf INFLATING, WARMING UP, \\AND PROBING THE PRE-BANGIAN UNIVERSE}
\\
\vspace*{1cm}
{\bf Gabriele VENEZIANO} \\
\vspace{0.3cm}
Theoretical Physics Division, CERN \\
CH - 1211 Geneva 23 \\
e-mail: {\tt venezia@nxth04.cern.ch}\\
\vspace*{2cm}
{\bf ABSTRACT} \\ \end{center}
\vspace*{5mm}

Classical and quantum gravitational instabilities, can, respectively, 
inflate and warm up a primordial  Universe satisfying
a  superstring-motivated principle of ``Asymptotic Past Triviality". 
A physically viable big bang is thus generated without invoking either large
fine-tunings or a long period of post-big bang inflation.
Properties of the pre-bangian Universe can be probed through 
its observable relics, which include:
i) a (possibly observable) stochastic gravitational-wave background;
ii) a (possible) new mechanism
for seeding the galactic magnetic fields; 
iii) a (possible) new source of large-scale
structure and CMB anisotropy.

\vspace*{1cm}

\vspace*{2.5cm}

\begin{flushleft}
CERN-TH/99-22 \\
February 1999
\end{flushleft}

\vfill\eject

\setcounter{page}{1}
\pagestyle{plain}

\section{INTRODUCTION}

I would like to begin this talk by asking a very simple question:
Did the Universe start ``small"?
The naive answer is: Yes, of course! However, a serious answer 
can only be given after defining the two keywords in the question: What do we mean by
``start"? and What is ``small" relative to? In order to be on the safe
side, let us take
the ``initial" time to be a bit larger than Planck's time,
$t_P  \sim 10^{-43} \; \rm{s}$. 
 Then, in standard Friedmann--Robertson--Walker (FRW) cosmology,
the initial size of the (presently observable)
Universe was about $10^{-2}~ \rm{cm}$. This is of course tiny w.r.t. its present size 
($\sim 10^{28}~ \rm{cm}$), yet huge w.r.t. the horizon at that time, i.e. 
w.r.t. $l_P = c t_P \sim 10^{-33}~ \rm{cm}$.
In other words, a few Planck times after the big bang,
 our observable Universe consisted 
of $(10^{30})^3 = 10^{90}$ Planckian-size, causally disconnected regions.

More precisely, soon after $t=t_P$, the Universe was characterized by a huge hierarchy
between its Hubble radius and  inverse temperature on one side, and its spatial-curvature
radius and  homogeneity scale on the other. The relative factor of (at least) $10^{30}$ appears
 as an incredible amount of fine-tuning on the initial state of the Universe,
corresponding to a huge asymmetry between time and space derivatives. Was this asymmetry
really there? And if so, can it be explained in any more natural way?

It is well known that a generic way to wash out inhomogeneities and spatial
curvature consists in introducing, in the history of the Universe, a long
period of accelerated expansion, called inflation \cite{KT}. 
This still leaves two alternative solutions: either the Universe was generic
at the big bang and became flat and smooth
because of a long {\it post}-bangian inflationary phase; 
 or it was
already flat and smooth at the big bang as the result of
 a long {\it pre}-bangian inflationary phase.

Assuming, dogmatically, that the Universe (and time itself) started at the big bang,
 leaves 
 only the first alternative. However, that solution has its own
problems, in particular those of 
fine-tuned initial conditions and inflaton potentials. Besides, it is quite difficult
to base standard inflation in the only known candidate theory of quantum gravity,
superstring theory. Rather, as we shall argue, superstring
 theory gives strong hints in favour of the
second (pre-big bang) possibility through two of its very basic properties, 
the first in relation
to its short-distance behaviour, the second from its modifications of General Relativity
even at large distance. Let us briefly comment on both.

\section {(Super)String inspiration}
\subsection{Short Distance}

Since the classical (Nambu--Goto)  action of a string 
is proportional to the area $A$ of the surface it sweeps, its quantization must introduce
a quantum of length $\lambda_s$ through:
\begin{equation}
S/\hbar =  A/\lambda_s^2\; .
\end{equation}
This fundamental length, replacing Planck's constant in quantum string theory \cite{GVFC},
plays the role of a minimal observable length, of an ultraviolet cut-off.
Thus, in string theory,  physical quantities are expected to be bound by appropriate
powers of $\lambda_s$, e.g. 
\begin{eqnarray}
H^2 \sim R \sim G\rho < \lambda_s^{-2} \nonumber \\
k_B T/\hbar < c \lambda_s^{-1} \nonumber \\
R_{comp} > \lambda_s \; .
\end{eqnarray}
 In other words, in quantum 
string theory (QST),  relativistic quantum mechanics should solve the singularity
problems in much  the same way as 
non-relativistic quantum mechanics solved the singularity
problem of the hydrogen atom by putting the electron and the proton a finite distance
apart. By the same token, QST gives us a rationale for asking  daring questions such as:
What was there before the big bang? Certainly, in no other present theory, can 
such a question
be meaningfully asked.

\subsection{Large Distance}

Even at large distance (low-energy, small curvatures), superstring theory does not
 automatically give Einstein's General Relativity. Rather, it leads to a 
scalar-tensor theory of the JBD variety. The new scalar particle/field $\phi$, 
the so-called dilaton, is unavoidable in string theory, and gets
reinterpreted as the radius of a new dimension of space in so-called M-theory \cite{M}.
By supersymmetry, the dilaton is massless to all orders in perturbation theory,
i.e. as long as supersymmetry remains unbroken. This raises the question:
Is the dilaton a problem or an opportunity? My answer is that it is possibly both;
and while we can try to avoid its potential dangers, we may try to use some of 
its properties to our advantage \dots~ Let me discuss how.

In string theory $\phi$ controls the strength of all forces, gravitational and gauge alike. 
One finds, typically:
\begin{equation}
 l_P^2 /\lambda_s^2  \sim \alpha_{gauge} \sim e^{\phi} \; ,
\label{VEV}
\end{equation}
showing the basic unification of all forces in string theory and the fact that, in 
our conventions, 
 the weak-coupling region coincides with $\phi \ll -1$.
In order not to contradict precision tests of the
Equivalence Principle and of the constancy of the gauge and gravitational
couplings in the recent past (possibly meaning several million years!)
we require \cite{TV} the dilaton to have a mass and to be frozen at the bottom of
 its own potential
{\it today}. This does not exclude, however, the possibility of the dilaton having
evolved cosmologically (after all the metric did!) within the weak coupling
region where it was practically massless. The amazing (yet simple) observation \cite{GV1}
is that, by so doing, the dilaton may have inflated the Universe!

A simplified argument, which, although not completely accurate, captures the essential
physical point, consists in writing the ($k=0$) Friedmann equation:
\begin{equation}
3 H^2 = 8 \pi G \rho\; ,
\end{equation}
 and in noticing that a growing dilaton (meaning through (\ref{VEV}) a growing $G$)
can drive the growth of $H$ even if the energy density of standard matter decreases
 in an expanding Universe. This new kind of inflation (characterized by 
growing $H$ and $\phi$) has been
termed dilaton-driven inflation (DDI). The basic idea of pre-big bang 
cosmology \cite{GV1,MG1,MG2}
is thus illustrated in Fig. 1: the dilaton  started at very
large negative values (where it is massless), ran over a potential hill, and finally reached,
sometime in our recent past, its final destination at the bottom of
its potential ($\phi = \phi_0$). Incidentally, as shown in Fig. 1, 
the dilaton of string theory can easily
roll-up ---rather than down--- potential hills, as a consequence of its non-standard
coupling to gravity.

\begin{figure}
\hglue 1.6 cm
 \epsfig{figure=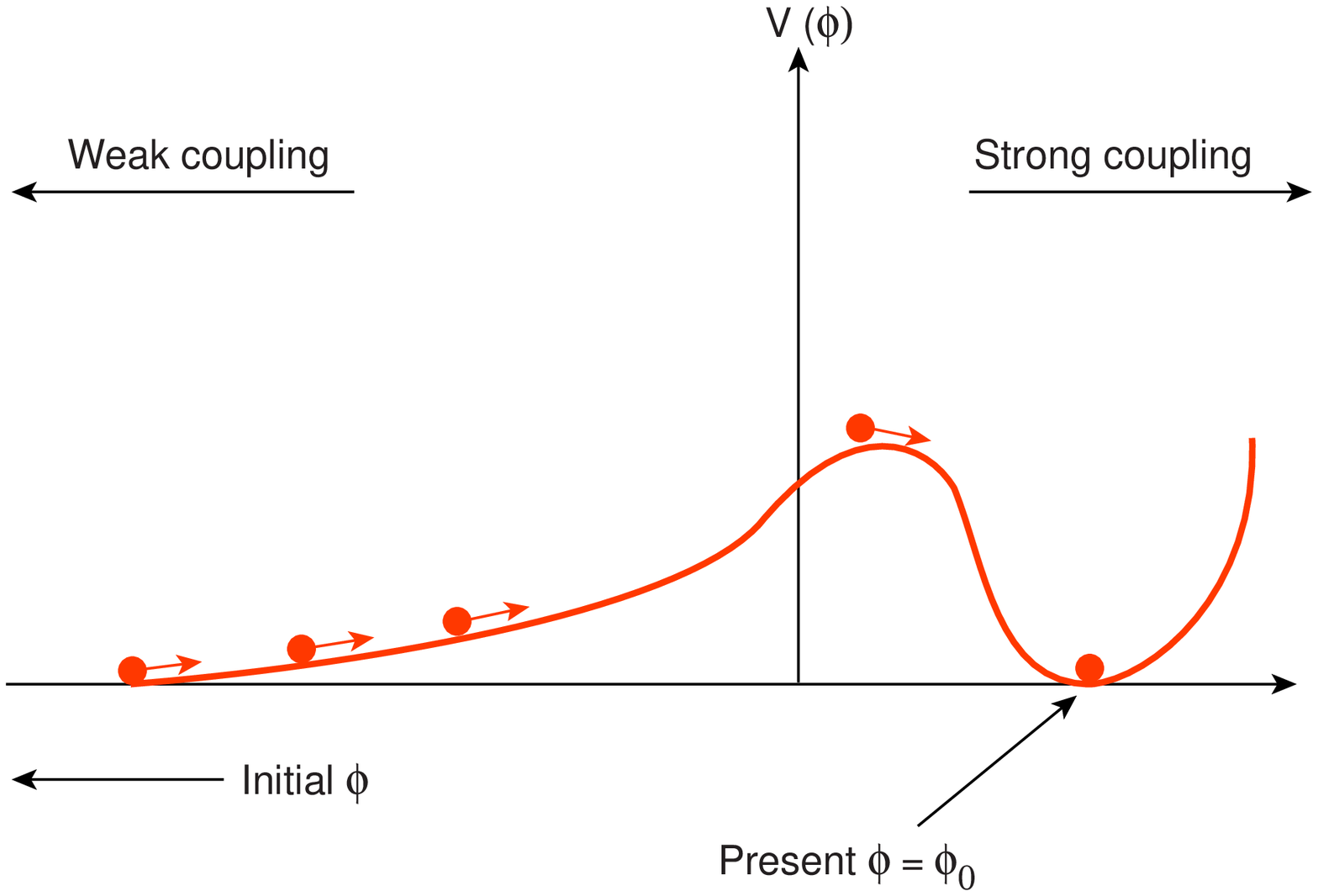,width=12cm}
 \caption[]{}
\end{figure}

DDI is not just possible. It exists as a class of (lowest-order) cosmological solutions
thanks to the duality symmetries of string cosmology. Under
a prototype example of these symmetries, the
so-called scale-factor duality \cite{GV1}, a FRW cosmology
evolving (at lowest order in derivatives) from a singularity in the past 
is mapped into a DDI cosmology going towards a singularity in the future. Of course, 
the lowest order approximation breaks down before either singularity is reached.
A (stringy) moment away from their respective
singularities, these two branches can easily be joined smoothly to give 
a single non-singular cosmology, at least mathematically. 
Leaving aside this issue for the moment (see Section V for more discussion),
let us go back to DDI.
 Since such a phase is characterized by growing coupling and curvature, it
 must itself have originated from
 a regime in which both quantities were very small. We take this as
 the main lesson/hint  to be
learned from low-energy string theory by raising  it to the level
of a new cosmological principle  \cite{BDV} of ``Asymptotic Past Triviality".

\section {Asymptotic Past Triviality}

The concept of Asymptotic Past Triviality (APT)
 is quite similar to that of ``Asymptotic Flatness",
familiar from General Relativity \cite{AF}. The main differences consist in making
 only  assumptions
concerning the asymptotic past (rather than future or space-like infinity)
 and in the additional presence
of the dilaton. It seems physically (and philosophically)
satisfactory to identify the beginning with simplicity (see e.g. entropy-related
 arguments
concerning the arrow of time). What could be  simpler than a trivial,
empty and flat Universe? Nothing of course! The problem is that such a Universe, 
besides being uninteresting, is also
non-generic. By contrast, asymptotically flat/trivial
 Universes are initially simple, yet generic in a precise mathematical sense.
Their definition involves exactly the right number of arbitrary ``integration constants"
(here full functions of three variables) to describe a general solution (one with
some general, qualitative features, though). This is why, by its very construction, 
this cosmology
cannot be easily dismissed as being fine-tuned.

It is useful to represent the situation in a Carter--Penrose diagram, 
 as  in Fig. 2. Here past infinity
consists of two pieces: time-like past infinity, which is shrunk to a point $I_-$, and
past null-infinity, $\cal{I}_-$ represented by a line at $45$ degrees. Note that this
region of the diagram is ``non-physical" in FRW cosmology, since it lies behind
 (i.e. before)
the big bang singularity (also shown in the diagram). Instead, we shall be giving
initial data infinitesimally close to 
$I_-$ and $\cal{I}_-$, and ask whether they will evolve in such a way as to generate
a physically interesting big bang-like state at some later time.
  Generating so much from so little looks a bit like a miracle. However,
 we will argue that it is precisely what
should be expected, owing to well-known classical and  quantum gravitational 
instabilities.

\begin{figure}
\hglue 5cm
 \epsfig{figure=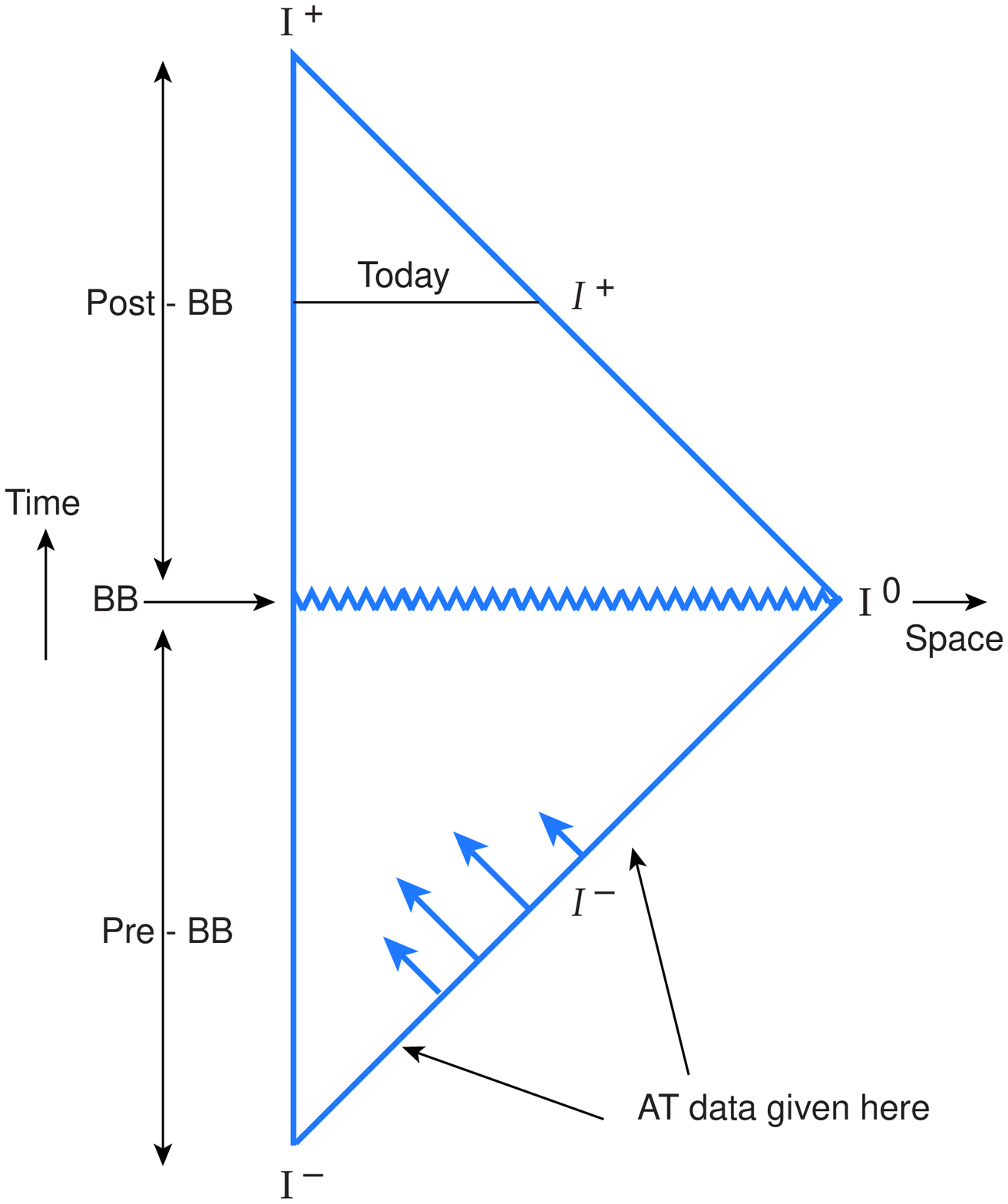,width=8cm}
 \caption[]{}
\end{figure}

\section {Inflation as a classical gravitational instability}

The assumption of APT  entitles us to treat the early history
of the Universe 
through the classical field equations of the low-energy
(because of the small curvature) tree-level (because of the weak coupling) 
effective action
of string theory. For simplicity, we will illustrate here the simplest case
of the gravi-dilaton system already compactified to
four space-time dimensions. Other fields and extra dimensions
will be mentioned below, when we discuss observable consequences.
The (string frame) effective action  then reads:
\begin{equation}
\Gamma_{eff} = \frac{1}{2 \lambda^{2}_s} \int d^4x \sqrt{-g}~ e^{-\phi}~
 ({\cal R}
+ \partial_\mu \phi \partial^\mu\phi) \; .
\label{leaction}
\end{equation}
In this frame, the string-length parameter $\lambda_s$ is a constant and the same is true of
the curvature scale at which we have to supplement eq. (\ref{leaction}) with corrections.
Similarly, string masses, when measured with the string metric, are fixed, while test strings 
sweep geodesic surfaces with respect to that metric.
 For all these reasons, even if we will allow metric
redefinitions in order to simplify our calculations, we shall eventually 
turn back to the string frame for the physical interpretation of the results. We stress, however,
that, while our intuition is not frame independent,
 physically measurable quantities are.

Even assuming APT, the problem of determining the properties of a generic
solution to the field equations implied by (\ref{leaction}) is a formidable one.
 Very luckily, however, we are  able to map our problem into one that has been much investigated,
both analytically and numerically, in the literature. This is done by going to the so-called
``Einstein frame". For our purposes, it simply amounts to the field redefinition
\begin{equation}
g_{\mu\nu} = g^{(E)}_{\mu\nu} e^{\phi - \phi_0} \; ,
\label{ESF}
\end{equation}
in terms of which (\ref{leaction}) becomes:
\begin{equation}
\Gamma_{eff} = \frac{1}{2 l^{2}_P} \int d^4x \sqrt{-g^{(E)}}
 ~~\left({\cal R}^{(E)}
- \half g_{(E)}^{\mu\nu} \partial_\mu \phi \partial_\nu\phi\right) \; ,
\label{EFaction}
\end{equation}
where $\phi_0$ ($l_P = \lambda_s e^{\phi_0/2}$) is the present value of
 the dilaton (of Planck's length). 

 Our problem is thus reduced to that of studying a massless scalar field
minimally coupled to gravity. Such a system has been considered by many authors,
in particular by Christodoulou \cite{Chr},
precisely in the regime of interest to us.  In line with the 
 APT postulate, in the analogue gravitational collapse problem,
 one assumes  very ``weak" initial data
with the aim of finding under which conditions  gravitational collapse later occurs.
Gravitational collapse  means that the (Einstein) metric (and the volume of 3-space)
 shrinks to zero at
a space-like singularity. However, typically,
the dilaton blows up at that same singularity. Given the relation (\ref{ESF})
between the Einstein and the (physical) string metric, we can easily imagine that
the latter blows up near the singularity as implied by DDI.

How generically does this happen? In this connection it is crucial to recall
the singularity theorems of Hawking and Penrose \cite{HP}, which state that, under some
general assumptions, singularities are inescapable in GR.
One can look at the validity of those assumptions in the case at hand and finds that
 all but one are automatically satisfied. The only condition to be
imposed is the existence of a closed trapped surface (a closed surface
 from where  future light cones lie entirely in the region inside the surface).
Rigorous results \cite{Chr} show that this condition cannot be waived: 
sufficiently weak initial data
do not lead to closed trapped surfaces, to collapse, or to singularities. Sufficiently
strong initial data do. But where is the border-line? This is not known in general,
but precise criteria do exist for particularly symmetric space-times, e.g. for those
endowed with spherical symmetry.
However, no matter what the general collapse/singularity criterion will 
eventually turn out to
be, we do know that:
\begin{itemize}
\item it cannot depend on an over-all additive constant in $\phi$;
\item it cannot depend on an over-all multiplicative factor in $g_{\mu\nu}$.
\end{itemize}
This is a simple consequence of the invariance (up to an over-all factor) of the effective
action (\ref{EFaction}) under shifts of the dilaton and rescaling of the metric (these 
properties depend crucially on the validity of the tree-level low-energy approximation
and on the absence of a cosmological constant). 

We conclude that, generically, some regions of space will undergo gravitational
collapse,  will form horizons and singularities therein, but nothing, at the level
of our approximations, will be able to fix either the size of the horizon
 or the value of $\phi$ at the onset of collapse. When this is translated 
into the string frame,
one is describing, in the region of space-time within the horizon, a period of DDI
in which both the initial value of the Hubble parameter and that of $\phi$ are left
arbitrary. These two initial parameters are very important, since they
determine the range of validity of our description. In fact, since both curvature and
coupling increase during  DDI, at some point the low-energy and/or tree-level
description is bound to break down. The smaller the initial Hubble parameter (i.e. the larger
the initial horizon size) and the smaller the initial coupling, the longer we can follow DDI
through the effective action equations and the larger the number of 
reliable e-folds that we shall gain.

This does answer, in my opinion, the objections raised recently \cite{TW} to the PBB
 scenario according to which it is
  fine-tuned. The situation here actually resembles that of
chaotic inflation \cite{chaotic}. Given some generic (though APT) initial data, we should ask
which is the distribution of sizes of the collapsing regions and of couplings therein.
Then, only the ``tails" of these distributions, i.e. those corresponding to
 sufficiently large,
and sufficiently weakly coupled regions, will produce Universes like ours, the rest will not.
The question of how likely a ``good" big bang is to take place is not very well posed
and can be greatly affected by anthropic considerations.

In conclusion, we may summarize recent progress on the problem of
initial conditions by saying that \cite{BDV}:
 {\bf Dilaton-driven inflation in string cosmology 
is as generic as gravitational collapse in General Relativity.}
At the same time,  having a sufficiently long period of 
DDI amounts to setting upper limits on two arbitrary
moduli of the classical solutions.

Our scenario is illustrated in Figs. 3 and 4, both taken
from Ref.\cite{BDV}. 
In Fig. 3, I show, for the spherically symmetric case,
 a Carter--Penrose diagram
 in which generic (but asymptotically trivial) dilatonic waves are
given around time-like ($I^-$) and null  ($\cal{I}^-$) past-infinity. In the
shaded region near $I^- , \cal{I}^-$, a weak-field solution holds. However, 
if a collapse criterion
is met, an apparent horizon, inside which a cosmological (generally inhomogeneous) PBB-like 
solution takes over, forms at some later
time. The future singularity of the PBB solution at $t=0$ is identified with the
space-like singularity of the black hole at $r=0$ (remember that $r$ is a 
time-like coordinate inside the horizon).
Figure 4 gives a $(2+1)$-dimensional sketch of a possible PBB Universe:  an original ``sea" of
dilatonic and gravity waves leads to collapsing regions of different initial size, possibly
to a scale-invariant distribution of them.
 Each one of these collapses is reinterpreted, in the string frame, as the process by which
 a baby Universe is born after a period of
PBB inflationary ``pregnancy", with the size of each baby Universe determined
 by the duration of 
its pregnancy, i.e. by the initial size of the 
corresponding collapsing region.  Regions initially 
larger than $10^{-13}~ {\rm cm}$ can generate Universes like ours,
smaller ones cannot.

\begin{figure}
\hglue 5cm
 \epsfig{figure=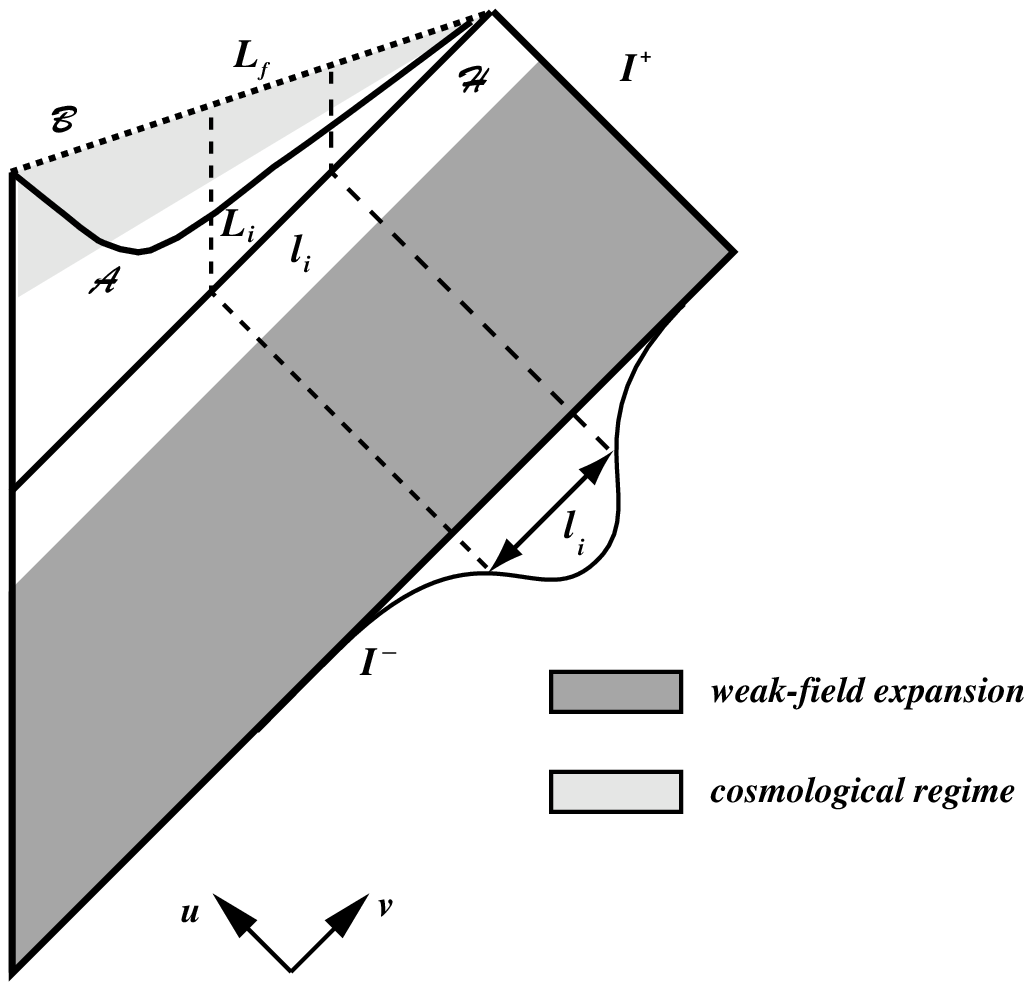,width=9cm}
 \caption[]{}
\end{figure}

\begin{figure}
\hglue 5cm
 \epsfig{figure=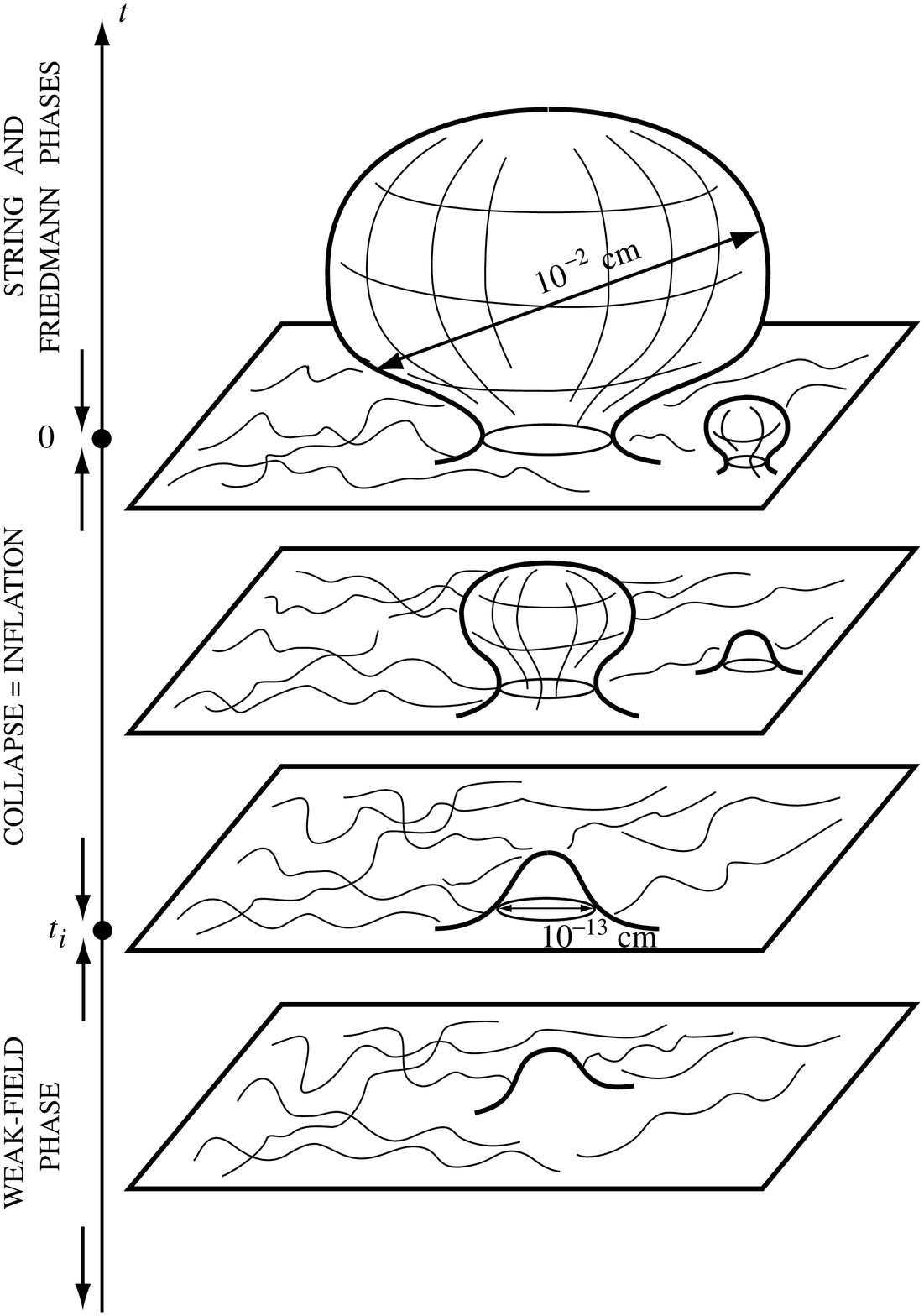,width=9cm}
 \caption[]{}
\end{figure}

A basic difference between the large numbers needed
in (non-inflationary) FRW cosmology and the large numbers needed in PBB cosmology
should be stressed at this point.
In the former, the ratio of two classical scales, e.g. of total curvature to its
spatial component, which is expected to be $O(1)$,
has to be taken as large as $10^{60}$. In the latter, the above ratio is 
initially $O(1)$ in the
collapsing/inflating region,
and ends up being very large in that region thanks to DDI.
However, the common order of magnitude of these two classical quantities 
is a free parameter, and is taken to be
much larger than a classically irrelevant quantum scale.

We can visualize analogies and differences between standard and pre-big bang inflation
by comparing Figs. 5a and 5b. In these, we sketch the evolution of the Hubble radius
and of a fixed comoving scale (here the one corresponding to the part of
the Universe presently
observable to us) as a function of time
in the two scenarios. The common feature is that the fixed comoving scale was
``inside the horizon" for some time during inflation, and possibly very deeply inside
at its onset. Also, in both cases, the Hubble radius at the beginning of inflation
had to be large in Planck units  and
the scale of homogeneity had to be at least as large. The difference between the two scenarios
is just in the behaviour of the Hubble radius during inflation: increasing in standard
inflation (a), decreasing in string cosmology (b). This is what makes PBB's ``wine glass" 
more elegant, and stable! Thus, while standard inflation is still facing the
 initial-singularity question and needs a non-adiabatic phenomenon to reheat the
Universe (a kind of small bang), PBB cosmology faces the singularity problem later,
combining it to the exit and heating problems (discussed in Sections
 V and VIB, respectively).

 In the end, what saves
PBB cosmology from fine-tuning is (not surprisingly!) supersymmetry. This is what
protects us from the appearance of a cosmological constant in the weak-coupling
regime. Even a relatively small cosmological constant would invalidate our
scale-invariance arguments and force DDI to be very short \cite{GV1}.
Thus, amusingly, while an effective cosmological constant is at the basis of
standard (post-big bang) inflation, its absence in the weak coupling region
is at the basis of PBB inflation. This may allow us to speculate that the 
absence (or extreme smallness) of the present cosmological constant
may be related to a mysterious degeneracy between the  perturbative
 and the  non-perturbative vacuum of superstring theory.

\section{The exit problem/conjecture}

We have argued that, generically, DDI, when studied at lowest
order in derivatives and coupling, evolves towards a singularity of the
big bang type. Similarly, at the same level of approximation, 
the non-inflationary solutions
emerge from a singularity. Matching these two branches in a smooth, non-singular way
has become known as the (graceful) exit problem in string cosmology \cite{exit}.
 It is, undoubtedly,
the most important theoretical problem facing the whole PBB scenario.

\begin{figure}
\hglue 5cm
 \epsfig{figure=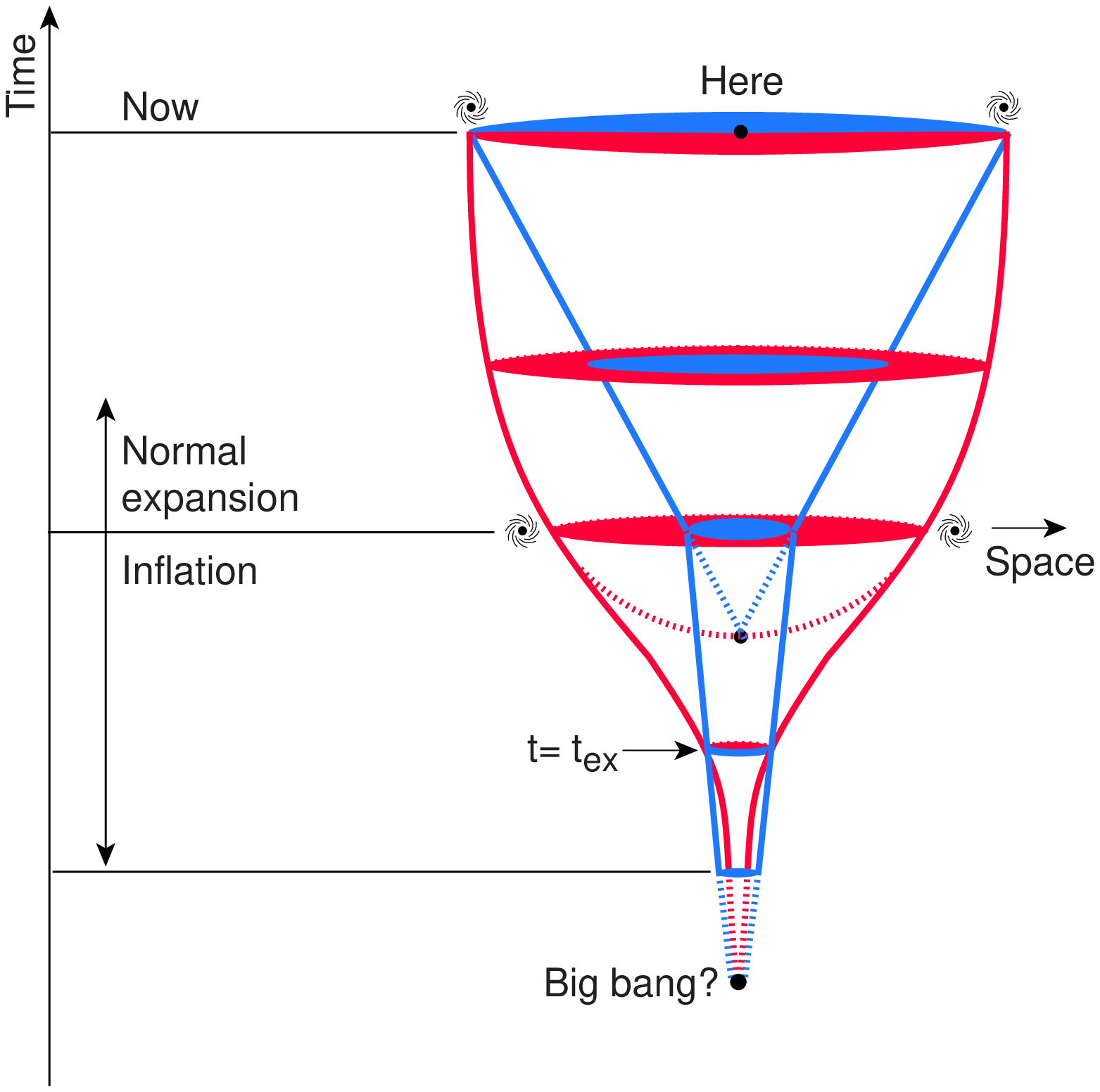,width=9cm}
\vglue 1 cm
\hglue 5cm
 \epsfig{figure=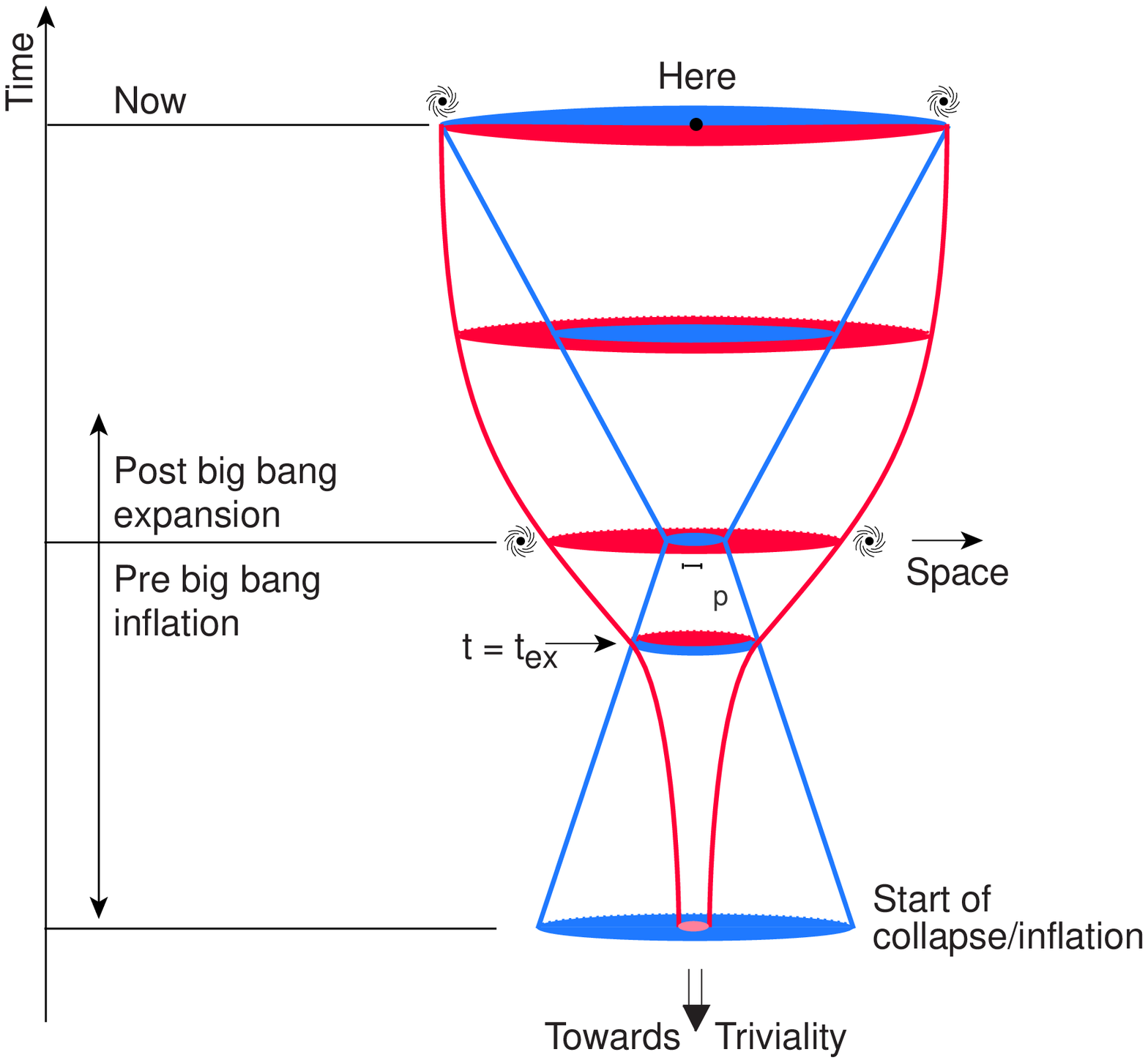,width=9cm}
 \caption[]{}
\end{figure}

There has been quite some progress recently on the exit problem. However,
for lack of space, I shall refer the reader to the literature \cite{exit} for details.
Generically speaking, toy examples have shown that DDI can flow, thanks to
higher-curvature corrections, into a de-Sitter-like phase, i.e. into a phase of
constant $H$ (curvature) and constant $\dot{\phi}$. This phase is expected to
last until loop corrections become important (see next section) and give rise to
a transition to a radiation-dominated phase. If these toy models serve as an indication,
the full exit problem can only be achieved at large coupling and curvature, a situation
that should be described by the newly invented M-theory \cite{M}.

It was recently pointed out \cite{MR} that the reverse order of events is also
possible. The coupling may become large {\it before} the curvature. In this case, at least
for some time, the low-energy limit of M-theory should be adequate: this limit is known \cite{M}
to give $D=11$ supergravity and is therefore amenable to reliable study. It is likely,
though not yet clear, that, also in this case, strong curvatures will 
have to be reached before the exit
can be completed. In the following, we will assume that:
\begin{itemize}
\item the big bang  singularity  is avoided thanks to the softness of string theory;
\item full exit to radiation occurs at strong coupling and curvature,
 according to a criterion
given in  Section VIB.
\end{itemize}

\section{ Observable relics and  heating the pre-bang Universe}
\subsection{PBB relics}
Since there are already  several review papers on this subject 
(e.g. \cite{GV95}),
 I will limit myself to mentioning the most recent developments, after recalling the basic
physical mechanism underlying particle production in cosmology \cite{quantum}.
A  cosmological (i.e. time-dependent)  background coupled to a given 
type of (small) inhomogeneous perturbation $\Psi$ enters 
the effective low-energy action in the form:
\begin{equation}
I ={\small\frac{1}{2}} \int d\eta\ d^3x\ S(\eta) \left[ \Psi
^{\prime 2}- (\nabla \Psi )^2\right].
\label{spertact}
\end{equation}
Here $\eta$ is the conformal-time coordinate, and a prime denotes
 $\partial/\partial\eta$. The function $S(\eta)$ (sometimes called
the ``pump" field)  is, for any given $\Psi$, a given function of the
scale factor $a(\eta)$, and of other scalar fields (four-dimensional 
dilaton $\phi(\eta)$,
moduli $b_i(\eta)$, etc.), which may appear non-trivially in the
background.

While it is clear that a constant pump field $S$ can be reabsorbed in a rescaling of $\Psi$, 
and is  thus ineffective, a time-dependent $S$ couples non-trivially to the fluctuation
and leads to the production of pairs of quanta (with equal and opposite momenta). 
One can
easily determine the pump fields for each one of the most interesting perturbations.
The result is:
\begin{eqnarray}
\rm{Gravity~waves,~dilaton}&:&   S = a^2 e^{-\phi} \nonumber \\
\rm{Heterotic~gauge~bosons}&:&  S =  e^{-\phi} \nonumber \\
\rm{Kalb-Ramond,~axions}&:&   S = a^{-2} e^{-\phi}\; .
\end{eqnarray}

A distinctive property of string cosmology is that the dilaton $\phi$ appears in some very
specific way in the pump fields. The consequences of this are very interesting:
\begin{itemize}

\item For gravitational waves and dilatons, the effect of $\phi$ is to slow down the behaviour
of $a$ (remember that both $a$ and $\phi$ grow in the pre-big bang phase). 
This is the reason why those spectra
are quite steep \cite{BGGV} and give small contributions at large scales.
Thus one of the most robust predictions of PBB cosmology is a small
tensor component in the CMB anisotropy\footnote{This, however,
refers just to first-order tensor perturbations; the mechanism ---described below--- of 
seeding CMB anisotropy through axions would also give a tensor (and a vector)
 contribution whose relative
magnitude is being computed.}.
The reverse is also true: at short scales, the expected yield in a stochastic background
 of gravitational waves is much larger than in standard inflationary
cosmology. This is easily understood: in standard inflation the GW spectrum
is either flat or slowly decreasing (as a function of frequency). Since COBE
 data \cite{COBE}
set a limit on the GW contribution at large scales, this bound holds a fortiori
at shorter scales, as those of interest for direct
GW detection. Thus, in standard inflation, one expects
\begin{equation}
\Omega_{GW} < 10^{-14}\; .
\end{equation}
Since the GW spectra of PBB cosmology are ``blue", the bound by COBE 
is automatically satisfied,
with no implication on the GW yield at interesting frequencies. Values of $\Omega_{GW}$ in
the range of
$10^{-6}$--$10^{-7}$ are possible in some regions of parameter space, which, according to
some estimates of sensitivities \cite{sens}, could be inside detection
capabilities in the near future. 

\item For gauge bosons there is no amplification of vacuum fluctuations in standard
cosmology, since a conformally flat metric (of 
the type forced upon by inflation) decouples from the
electromagnetic (EM) field precisely in $D=3+1$ dimensions. 
As a very general remark, apart from pathological
solutions, the only background field which, through its cosmological variation,
can amplify EM (more generally gauge-field) quantum fluctuations is the effective gauge coupling
itself \cite{Ratra}. By its very nature,
 in the pre-big bang scenario  the effective gauge coupling inflates 
together with space during the PBB phase.
It is thus automatic that any efficient PBB inflation brings together
 a huge variation of the effective gauge coupling and
thus a very large amplification of the primordial
EM fluctuations \cite{GGV,BMUV2,BH}. This can possibly provide the long-sought origin 
for the primordial seeds of the observed
galactic magnetic fields. 
Notice, however, that, unlike GW, EM perturbations interact quite
considerably with the hot plasma of the early (post-big bang) Universe.
Thus, converting the primordial seeds into those that may have existed
at the proto-galaxy formation epoch is by no means a trivial exercise.
Work is in progress to try to adapt existing codes \cite{Ostreiker} to the evolution
of our primordial seeds.
\item Finally, for Kalb--Ramond fields and axions, $a$ and $\phi$ work in the same direction and
spectra can be large even at large scales \cite{Copeland}.
An interesting fact is that, unlike the GW spectrum, that of axions is very sensitive to
the cosmological behaviour of internal dimensions during the DDI epoch.
On one side, this makes  the model less predictive. On the other, it tells us that
axions represent a window over the multidimensional cosmology expected generically
from string theories, which must live in more that four dimensions.
Curiously enough, the axion spectrum becomes exactly HZ (i.e. scale-invariant) when
all the nine spatial dimensions of superstring theory evolve in a rather
symmetric way \cite{BMUV2}.
In situations near this particularly symmetric one, axions are able to provide
a new mechanism for generating large-scale CMB anisotropy and LSS.

A recent calculation \cite{Durrer} of the effect gives, for massless axions, 
\begin{equation}
l(l+1) C_l \sim O(1) \left({H_{max}\over M_P}\right)^4 (\eta_0 k_{max})^{-2\alpha} 
{\Gamma(l+\alpha) \over \Gamma(l- \alpha)}\; , 
\end{equation}
where  $C_l$ are the usual  coefficients of the multipole expansion of $\Delta T/T$
\begin{equation}
\langle \Delta T/T(\vec{n})~~ \Delta T/T(\vec{n}')\rangle ~ = ~
 \sum_l (2l+1) C_l P_l(\cos\theta)\; ,
\end{equation}
and the parameters $H_{max}, k_{max}, \alpha$ are defined by the primordial
axion energy spectrum in critical units as:
\begin{equation}
\Omega_{ax}(k) = \left({H_{max}\over M_P}\right)^2 (k/ k_{max})^{\alpha} \; .
\end{equation}

In string theory, as repeatedly mentioned, we expect $H_{max}/ M_P \sim M_s/M_P \sim 1/10$
and $\eta_0 k_{max} \sim 10^{30}$, while the exponent $\alpha$ depends on the explicit
PBB background with the above-mentioned HZ case corresponding to $\alpha =0$. The standard
tilt parameter $n = n_s$ ($s$ for scalar) is given by $n = 1 + 2 \alpha$ and is found, by COBE, 
to lie between $0.9$ and $1.5$, corresponding to $0 < \alpha < 0.25$ (a negative $\alpha$ leads
to some theoretical problems). With these inputs we can see that the correct normalization
($C_2 \sim 10^{-10}$) is reached for $\alpha \sim 0.2$, which is just in the 
middle of the allowed range. In other words, unlike in standard inflation, we cannot
predict the tilt, but when this is given, we can predict (again unlike in standard inflation)
the normalization.

Our model, being of the isocurvature type, bears some resemblance
 to the one recently advocated by Peebles \cite{Peebles}
and, like his, is expected to contain some calculable amount
 of non-Gaussianity, which is being 
calculated and will be checked by the future satellite measurements (MAP, PLANCK).
\item Many other perturbations, which arise in generic compactifications of superstrings,
 have also
been studied, and lead to interesting spectra. For lack of time, I will refer to
the existing literature \cite{BMUV2,BH}.
\end{itemize}

\subsection{Heat and entropy as a quantum gravitational instability }

Before closing this section, I wish to recall how one sees the very origin of the hot big bang
in this scenario. One can easily estimate the total energy stored in the
quantum fluctuations, which were amplified by the pre-big bang backgrounds. 
The result is, roughly,
\begin{equation}
\rho_{quantum} \sim N_{eff} ~ H^4_{max} \; ,
\label{rhoq}
\end{equation}
where $N_{eff}$ is the effective number of species that are amplified and $H_{max}$ is the maximal
curvature scale reached around $t=0$. We have already argued that $H_{max} \sim M_s =
 \lambda_s^{-1}$,
 and we know that, 
in heterotic string theory, $N_{eff}$ is in the hundreds. Yet this rather huge energy density
is very far from critical, as long as the dilaton is still in the weak-coupling region, 
justifying our neglect of back-reaction effects. It is very tempting to assume \cite{BMUV2} that,
precisely when the dilaton reaches a value such that $\rho_{quantum}$ is critical, the Universe
will enter the radiation-dominated phase. This PBBB (PBB bootstrap) constraint gives, typically:
\begin{equation}
e^{\phi_{exit}}  \sim 1/N_{eff}\;\; ,
\label{PBBB}
\end{equation}
i.e. a value for the dilaton close to its present value.

The entropy in these quantum fluctuations can also be estimated following
some general results \cite{entropy}. The result for the density of entropy $S$ is, as expected 
\begin{equation}
S \sim N_{eff} H_{max}^3\;.
\end{equation}
It is easy to check that, at the assumed time of exit given by (\ref{PBBB}), 
this entropy saturates a 
recently proposed holography bound \cite{FS}. This also turns out to be a physically
acceptable value for the entropy of the Universe just after the big bang: a large entropy
on the one hand (about $10^{90}$); a small entropy for the total mass and size of the observable
Universe on the other, as often pointed
out by Penrose \cite{PenEntr}. Thus, PBB cosmology neatly explains why the
 Universe, at the big bang,
looks so fine-tuned (without being so) and provides a natural arrow of time in the direction
of higher entropy.

\section{Conclusions}
\begin{itemize}
\item
Pre-big bang (PBB) cosmology is a ``top--down" rather than a ``bottom--up" approach to cosmology.
This should not be forgotten when testing its predictions. 
\item
It does not need to invent an inflaton, or to fine-tune its potential; inflation is
``natural" thanks to the duality symmetries of string cosmology.
\item
It makes use of a classical gravitational instability to inflate the Universe,
and of a quantum instability to warm it up.
\item
The problem of initial conditions  ``decouples" from the singularity problem; 
it is classical, scale-free, and unambiguously defined. Issues of fine tuning
can be addressed and, I believe, answered.
\item The spectrum of large-scale perturbations has become more promising 
through the invisible axion of string theory, while the possibility of explaining the seeds of
galactic magnetic fields remains a unique prediction of the model.
\item The main conceptual (technical?) problem remains that of providing
a fully convincing mechanism for (and a detailed description of) the
pre-to-post-big bang transition. It is very likely that such a mechanism will involve both high
curvatures and large coupling and should therefore be discussed  in the (yet to be
fully constructed) M-theory \cite{M}. New ideas borrowed from
such theory and from  D-branes \cite{branes,MR} could help in this respect.
\item
Once/if this problem will be solved, predictions will become more precise and robust, but,
even now, with some mild assumptions, several tests are (or will soon become) possible, e.g.
\begin{itemize}
\item the tensor contribution to $\Delta T/T$ should be very small 
(see, however, footnote Section VI);
\item some non-Gaussianity in $\Delta T/T$ correlations is expected, and calculable.
\item the axion-seed mechanism should lead to
 a characteristic acoustic-peak structure, which is being calculated;
\item it should be possible to convert 
the predicted seed magnetic fields into observables by
using some reliable code for their late evolution;
\item a characteristic spectrum of stochastic gravitational waves is expected to surround us
and could be large enough to be measurable within a decade or so.
\end{itemize}
\end{itemize}


\begin{thebibliography}{99}
\bibitem{KT} E. W. Kolb and M. S. Turner, {\em The Early Universe} 
(Addison-Wesley, Redwood City, CA, 1990); 
A.D. Linde, {\em Particle Physics and Inflationary Cosmology} 
(Harwood, New York, 1990). 
\bibitem{GVFC} G. Veneziano, Europhys. Lett. {\bf 2} (1986) 133; 
{\em The Challenging Questions}, Erice, 1989,
 ed. A. Zichichi (Plenum Press, New York, 1990), p. 199.
\bibitem{M} See, e.g., E. Witten, Nucl. Phys. {\bf B443} (1995) 85;\\
 P. Horawa and E. Witten, Nucl. Phys. {\bf B460} (1996) 506.
\bibitem{TV} T.R. Taylor and G. Veneziano, Phys. Lett. {\bf B213} (1988) 459.
\bibitem{GV1} G. Veneziano, Phys. Lett. {\bf B265}  (1991) 287.
\bibitem{MG1}
 M. Gasperini and G. Veneziano, Astropart. Phys. {\bf  
1} (1993)  317, Mod. Phys. Lett. {\bf A8}  (1993) 3701, Phys. Rev. {\bf D50} (1994)  
2519.
\bibitem{MG2} An  updated collection of papers on the PBB
scenario is available at {\rm http}://{\rm www.to.infn.it/\~{}gasperin/}.
\bibitem{BDV} A. Buonanno, T. Damour and G. Veneziano, {\em Pre-big 
bang bubbles from the gravitational instability of generic
string vacua}, hep-th/9806230; 
see also, G. Veneziano,  Phys. Lett. {\bf B406 } (1997) 297;
 A. Buonanno, K.A. Meissner,  C. Ungarelli 
and G. Veneziano, Phys. Rev. {\bf D57} (1998) 2543, and references therein.
\bibitem{AF} R. Penrose, {\em Structure of space-time}, in {\em Battelle Rencontres},
ed. C. Dewitt and  \\ J.A. Wheeler, Benjamin, New York, 1968.
\bibitem{Chr} D. Christodoulou, Commun. Pure Appl. Math. {\bf 56} (1993) 1131, 
and references therein.
\bibitem{HP} R. Penrose, Phys. Rev. Lett. {\bf 14} (1965) 57;
S. W. Hawking and R. Penrose, Proc. Roy. Soc. Lond. {\bf A314} (1970) 529.
\bibitem{TW} M. Turner and E. Weinberg, Phys. Rev. {\bf D56} (1997) 4604; 
  N. Kaloper, A. Linde and  R. Bousso, {\em Pre-big bang 
requires the Universe to be
exponentially large from the very beginning}, hep-th/9801073.
\bibitem{chaotic} A. Linde, Phys. Lett. {\bf 129B} (1983) 177.
\bibitem{exit} R. Brustein and G. Veneziano, Phys. Lett. {\bf B329} (1994) 429; 
N. Kaloper, R. Madden and K.A. Olive, Nucl. Phys. {\bf B452} (1995) 677, 
Phys. Lett. {\bf B371} (1996) 34; R. Easther, K. Maeda and D. Wands, Phys. Rev.
 {\bf D53} (1996) 4247;
M. Gasperini, M. Maggiore and G. Veneziano,  
 Nucl. Phys. {\bf B494} (1997) 315;
R. Brustein and R. Madden,  Phys. Lett. {\bf B410} (1997) 110,
 Phys. Rev. {\bf D57} (1998) 712.
\bibitem{MR} M. Maggiore and A. Riotto, {\em D-branes and Cosmology}, hep-th/9811089;\\
see also T. Banks, W. Fishler and L. Motl, {\em Duality versus Singularities},
hep-th/9811194.
\bibitem{GV95} G. Veneziano,  in {\em 
String Gravity and Physics at the Planck Energy Scale}, Erice, 1995, eds. 
N. Sanchez and A. Zichichi  (Kluver Academic Publishers, Boston, 1996), p. 285;
 M. Gasperini, ibid., p. 305. 
\bibitem{quantum} See, e.g., V. F. Mukhanov, A. H. Feldman and R. H. Brandenberger, 
 Phys. Rep.  {\bf 215} (1992) 203.
\bibitem {BGGV}
R. Brustein, M. Gasperini, M. Giovannini and G. Veneziano, Phys. Lett. {\bf B361}  
(1995) 45;  R. Brustein et al., Phys. Rev. {\bf D51} (1995) 6744.
\bibitem{COBE} G. F. Smoot et al.,  Ap. J. {\bf 396} (1992) L1;
 C. L. Bennet et al.,  Ap. J. {\bf 430}  (1994) 423.
\bibitem{sens}
P. Astone et al., Phys. Lett. {\bf B385} (1996) 421;
 B. Allen and R. Brustein, Phys. Rev. {\bf D55} (1997) 970.
\bibitem{Ratra} B. Ratra, Astrophys. J. Lett. {\bf 391} (1992) L1.
\bibitem{GGV}
M. Gasperini, M. Giovannini and G. Veneziano, Phys. Rev. Lett.  {\bf 75} (1995) 3796;\\ 
D. Lemoine and M. Lemoine, Phys. Rev. {\bf D52} (1995) 1955.
\bibitem{BMUV2}
A. Buonanno, K. A. Meissner, C. Ungarelli and G. Veneziano, 
JHEP {\bf 1} (1998) 4; 
\bibitem{BH} 
R. Brustein and M. Hadad, Phys. Rev. {\bf D57} (1998) 725.
\bibitem{Ostreiker} R. M. Kulsrud, R. Cen, J. P. Ostriker and
 D. Ryu, Ap. J. {\bf 480} (1997) 481.
\bibitem{Copeland} E.J. Copeland, R. Easther and D. Wands,
 Phys. Rev. {D56} (1997) 874;
E.J. Copeland, J.E.  Lidsey and D. Wands,
Nucl. Phys. {\bf B506} (1997) 407.
\bibitem{Durrer} R. Durrer, M. Gasperini, M. Sakellariadou and G.
Veneziano, Phys. Lett. {\bf B436} (1998) 66, Phys. Rev. {D59} (1999) 043511;
M. Gasperini and G. Veneziano, Phys. Rev. {D59} (1999) 043503.
\bibitem{Peebles} P. J. E. Peebles, {\em An isocurvature CDM cosmogony. I and II},
astro-ph/9805194, and  astro-ph/9805212.
\bibitem{entropy} M. Gasperini and M. Giovannini, Phys. Lett. {\bf B301} (1993) 334;
Class. Quant. Grav. {\bf 10} (1993) L133;
 R. Brandenberger, V. Mukhanov and T. Prokopec,  Phys.
Rev. Lett. {\bf 69} (1992)  3606;  Phys. Rev.  {\bf D48} (1993) 2443.
\bibitem{FS} W. Fischler and L. Susskind, {\em Holography and Cosmology}, hep-th/9806039;
see also \\ D. Bak and S.-J. Rey, {\em Holographic principle and string cosmology},
hep-th/9811008;\\ A. K. Biswas, J. Maharana and R.K. Pradhan, {\em The holography
principle and pre-big bang cosmology}, hep-th/9811051.
\bibitem{PenEntr} see, e.g., R. Penrose, {\em The Emperor's new mind}, 
(Oxford University Press, New York, 1989), Chapter 7.
\bibitem{branes} A. Lukas, B.A. Ovrut and D. Waldram, Phys. Lett. {\bf B393} (1997) 65; 
Nucl. Phys. {\bf B495} (1997) 365;
F. Larsen and F. Wilczek, Phys. Rev. {\bf D55} (1997) 4591;
N. Kaloper, Phys. Rev. {\bf D55} (1997) 3394; 
H. Lu, S. Mukherji and C.N. Pope, Phys. Rev. {\bf D55} (1997) 7926;
 R. Poppe and S. Schwager, Phys. Lett.  
{\bf B393} (1997) 51;
A. Lukas and B. A. Ovrut, Phys. Lett. {\bf B437} (1998) 291;
N. Kaloper, I. Kogan and K. A. Olive, Phys. Rev. {\bf D57} (1998) 7340.
\end{thebibliography}
\end{document}